\documentstyle[12pt,prd,aps,epsfig,preprint]{revtex}

\begin{document}
\title{{\bf Series Solutions of the }${\bf N}${\bf -Dimensional Position-Dependent
Mass Schr\"{o}dinger Equation with a General Class of Potentials }}
\author{Sameer M. Ikhdair\thanks{%
sikhdair@neu.edu.tr} and \ Ramazan Sever\thanks{%
sever@metu.edu.tr}}
\address{$^{\ast }$Department of Physics, \ Near East University, Nicosia, North
Cyprus, Mersin-10, Turkey\\
$^{\dagger }$Department of Physics, Middle East Technical \ University,
06531 Ankara, Turkey.}
\date{\today
}
\maketitle

\begin{abstract}
The analytical solutions of the $N$-dimensional Schr\"{o}dinger equation
with position-dependent mass for a general class of central potentials is
obtained via the series expansion method. The position-dependent mass is
expanded in series about origin. As a special case, the analytical
bound-state series solutions and the recursion relation of the
linear-plus-Coulomb (Cornell) potential with the decaying position-dependent
mass $m=m_{0}e^{-\lambda r}$ are also found.


PACS\ number: 03.65.-w, 03.65.Ca, 03.65.Ge, 02.30.Hq
\end{abstract}


\section{Introduction}

\noindent The solution of the Schr\"{o}dinger equation with
position-dependent mass for any spherically symmetric potential has
attracted attention over the past years [1-17]. The motivation in this
direction arises from considerable applications in the different fields of
the material science and condensed matter physics. For instance, such
applications in the case of the bound states in quantum system [4], the
nonrelativistic Green's function for quantum systems with the
position-dependent mass [5], the Dirac equation with position-dependent mass
in the Coulomb field [5], electronic properties of semiconductors [10], $%
^{3}He$ cluster [11], quantum dots [12], semiconductor heterostructures
[13,14], quantum liquids [15], the dependence of energy gap on magnetic
field in semiconductor nano-scale quantum rings [16], the solid state
problems with the Dirac equation [17]. Almost all of those works mentioned
above were focused on obtaining the energy eigenvalues and the potential
function for the given quantum system with the position-dependent mass. The
wave functions were either obtained by the solutions to the Schr\"{o}dinger
equation with the constant mass, or a few lower excited states were obtained
by acting of the creation operator on the ground state. The effective
potentials are the sum of the real potential form and the modification terms
emerged from the location dependence of the effective mass [2]. Taking into
consideration the position-dependent mass, the aim of this work is to carry
out the analytical solutions of the $N-$dimensional Schr\"{o}dinger equation
with position-dependent mass for a general class of static quarkonium
potentials by the series expansion method used in [18,19]. Additionally, we
investigate the linear-plus-Coulomb (Cornell) potential case [20,21].

The contents of this paper is as follows. In Section \ref{TND}, we present
the $N$ dimensional Schr\"{o}dinger equation with position-dependent mass
for any spherically symmetric potential. In Section \ref{TSS} the analytical
bound-state series solutions of a general class of quarkonium potentials. In
Section \ref{TAS}, we study the analytical series solutions of the Cornell
potential for a particle of an exponentially decaying mass, $%
m=m_{0}e^{-\lambda r},$ $\lambda >0,$ case and then give an example for the
analytical calculations of the Coulomb's wave function. Finally, in Section
\ref{CR} we give our concluding remarks.

\section{The $N$-Dimensional Position-Dependent Mass Schr\"{o}dinger Equation%
}

\label{TND}The wave Schr\"{o}dinger equation with position-dependent mass
for a spherically symmetric potential $V(r)$ in $N$-dimensional space (in $%
\hbar =1$ units):

\begin{equation}
{\bf \nabla }_{N}\frac{1}{m}{\bf \nabla }_{N\text{ }}\psi (r)+2\left[ E-V(r)%
\right] \psi (r)=0,
\end{equation}
where the wave function is defined by [20,21,22,23]
\begin{equation}
\psi (r)=e^{-(N-1)/2}R_{n,l}(r)Y_{l,m}(x).
\end{equation}
and $m=m(r).$ We substitute ${\bf \nabla }_{N}\frac{1}{m(r)}{\bf \nabla }_{N%
\text{ }}\psi (r)=\left( {\bf \nabla }_{N}\frac{1}{m(r)}\right) \cdot \left(
{\bf \nabla }_{N\text{ }}\psi (r)\right) +\frac{1}{m(r)}{\bf \nabla }_{N%
\text{ }}^{2}\psi (r)$ into Eq.(1) and obtain the $N-$dimensional
position-dependent mass radial Schr\"{o}dinger equation

\begin{equation}
\left\{ \frac{d^{2}}{dr^{2}}+\frac{m^{\prime }}{m}\left( \frac{N-1}{2r}-%
\frac{d}{dr}\right) -\frac{\left[ k-1\right] \left[ k-3\right] }{4r^{2}}%
+2m(r)\left[ E-V(r)\right] \right\} R_{n,l}(r)=0,
\end{equation}
where $k=N+2l$ and $m^{\prime }(r)=dm/dr.$ It can be clearly seen that for $%
m^{\prime }=0$ case$,$ the above equation reduces to the well known equation
with constant mass used in Refs.[20,21,22,23]. In the present work, we are
concerned in bound states, i.e., $E<0.$ On the other hand, one should be
careful about the behavior of the wave function $R(r)$ near $r=0$ and $%
r\rightarrow \infty .$ It may be mentioned that $R(r)$ behaves like $%
r^{(k-1)/2}$ near $r=0$ and it should be normalizable. We choose the wave
function [18,19,24,25]

\begin{equation}
R_{n,l}(r)=r^{(k-1)/2}e^{-br}u(r),\text{ \ }b=\sqrt{-2m_{0}E},\text{\ \ }
\end{equation}
where $m_{0}$ is the initial value of mass. It should be noted that $N$ and $%
l$ enter into expression (3) in the form of the combination $k=N+2l.$
Consequently, the solutions for a particular central potential $V(r)$ are
the same as long as $k$ remains unaltered. Thus, the $s$-wave eigensolution $%
(R_{n0})$ and eigenvalues $(E)$ in four-dimensional space are identical to
the $p$-wave two-dimensional solutions. By substituting Eq.(4) into Eq.(3),
we obtain

\[
\left[ (k-1)\frac{d}{dr}-(k-1)b-\frac{m^{\prime }}{m}l\right] u(r)+
\]
\begin{equation}
\left[ \frac{d^{2}u(r)}{dr^{2}}-\left( 2b+\frac{m^{\prime }}{m}\right) \frac{%
du(r)}{dr}+\frac{m^{\prime }}{m}bu(r)-2mV(r)u(r)+2E(m-m_{0})u(r)\right] r=0.
\end{equation}
Next we consider a series solutions for the above reduced radial wave
Schr\"{o}dinger equati\i n.

\section{The Series Solution with a Class of Static Potentials}

\label{TSS}Equation (3) cannot be solved exactly except for some particular
cases. Neverthless, we can get an approximate solution using the series
expansion method [18,19]. In application, we consider here a group of
central potentials belong to the following general form [23]

\begin{equation}
V(r)=-V_{1}r^{-\alpha }+V_{2}r^{\beta }+V_{3},
\end{equation}
where $V_{1}$ and $V_{2}$ are positive coupling constants whereas the
constant $V_{3}$ may be of either sign. Moreover, this group of potentials
satisfies the boundary conditions stated in [20,21,22,23]. This class of
generality for potentials (6) is used to produce the bound state energy
spectra for quarkonium systems [20,21,22,23]. It comprises a well-known
potential, e.g., the Cornell potential (we set $\alpha =\beta =1,$ $V_{1}=A=%
\frac{4}{3}\alpha _{s},$ $V_{2}=B,$ $V_{3}=C)$ (cf. Ref.[21]).

Now we try the following series expansions about origin:

\begin{equation}
m(r)=\sum_{\nu =0}^{\infty }b_{\nu }r^{\nu }=b_{0}+\sum_{\nu =1}^{\infty
}b_{\nu }r^{\nu },\text{ \ }b_{0}=m_{0}
\end{equation}
\begin{equation}
\frac{m^{\prime }}{m}=\sum_{\nu =0}^{\infty }b_{\nu }^{\prime }r^{\nu },
\end{equation}
together with the series expansion for $u(r)$

\begin{equation}
u(r)=\sum_{i=0}^{\infty }a_{i}r^{i},\text{ \ }a_{0}\neq 0
\end{equation}
and substitute Eqs.(6)-(9) into Eq.(5), we obtain the following relation

\[
\sum_{i=0}^{\infty }\left[
-(k-1)ba_{i}r^{i}-2bia_{i}r^{i}+(k-1)ia_{i}r^{i-1}+i(i-1)a_{i}r^{i-1}\right.
\]
\[
-\sum_{\nu =0}^{\infty }b_{\nu }^{\prime }\left( ia_{i}r^{\nu
+i}+la_{i}r^{\nu +i}-ba_{i}r^{\nu +i+1}\right)
+b^{2}a_{i}r^{i+1}+2E\sum_{\nu =0}^{\infty }b_{\nu }a_{i}r^{\nu +i+1}
\]
\begin{equation}
\left. +2V_{1}\sum_{\nu =0}^{\infty }b_{\nu }a_{i}r^{\nu +i-\alpha
+1}-2V_{2}\sum_{\nu =0}^{\infty }b_{\nu }a_{i}r^{\nu +i+\beta
+1}-2V_{3}\sum_{\nu =0}^{\infty }b_{\nu }a_{i}r^{\nu +i+1}\right] =0,
\end{equation}
with $b^{2}=-2m_{0}E.$ On the other hand, we define

\begin{equation}
M_{i}=\sum_{j,\nu =0}^{j+\nu =i}a_{j}b_{\nu };\text{ \ }i\geq 0
\end{equation}
\begin{equation}
M_{i}^{\prime }=\sum_{j,\nu =0}^{j+\nu =i}a_{j}b_{\nu }^{\prime };\text{ \ }%
i\geq 0
\end{equation}
\begin{equation}
T_{i}=\sum_{j,\nu =0}^{j+\nu =i}ja_{j}b_{\nu }^{\prime };\text{ \ }i\geq 1,%
\text{ \ }T_{0}=0.
\end{equation}
Setting the coefficients of the power of $r^{n}$ to be zero, we obtain the
following recurrence relation of the bound energy spectrum

\[
n(n+1)a_{n+1}+(k-1)(n+1)a_{n+1}-(k-1)ba_{n}-2bna_{n}-lM_{n}^{\prime
}+bM_{n-1}^{\prime }-T_{n}
\]

\begin{equation}
+2EM_{n-1}+b^{2}a_{n-1}+2V_{1}M_{n+\alpha -1}-2V_{2}M_{n-\beta
-1}-2V_{3}M_{n-1}=0,
\end{equation}
with the final radial wave functions

\begin{equation}
R_{n,l}(r)=r^{(k-1)/2}e^{-br}\sum_{i=0}^{\infty }a_{i}r^{i},\text{ \ \ \ }%
a_{0}\neq 0.
\end{equation}
We now present some special cases.

\subsection{Coulomb Potential}

For the solutions of the Coulomb problem in $N$-dimensional space, we set $%
V_{1}=Z,$ $V_{2}=V_{3}=0,$ $\alpha =1,$ $\beta =0,$ then we get its
recurrence relation from Eq.(14):

\[
n(n+1)a_{n+1}+(k-1)(n+1)a_{n+1}-(k-1)ba_{n}-2bna_{n}
\]

\begin{equation}
-lM_{n}^{\prime }+bM_{n-1}^{\prime }-T_{n}+2EM_{n-1}+b^{2}a_{n-1}+2ZM_{n}=0,
\end{equation}
with the radial wave functions are given in Eq.(15), cf. Ref.[24]. This case
was treated in Ref..[19].

\subsection{Harmonic Oscillator Potential}

For the solutions of the harmonic oscillator problem in $N$-dimensional
space, we set $V_{2}=\omega ^{2},$ $V_{1}=V_{3}=0,$ $\alpha =0,$ $\beta =2,$
then the recurrence relation (14) becomes

\[
n(n+1)a_{n+1}+(k-1)(n+1)a_{n+1}-(k-1)ba_{n}-2bna_{n}
\]

\begin{equation}
-lM_{n}^{\prime }+bM_{n-1}^{\prime }-T_{n}+2EM_{n-1}+b^{2}a_{n-1}-2\omega
^{2}M_{n-3}=0,
\end{equation}
with the radial wave functions are given in Eq.(15). This case was also
treated in Ref.[19]..

\subsection{Confining Linear Potential}

For the solutions of the confining linear potential in $N$-dimensional
space, we set $V_{1}=V_{3}=0,$ $V_{2}=B,$ $\alpha =0,$ $\beta =1,$ then
Eq.(14) becomes

\[
n(n+1)a_{n+1}+(k-1)(n+1)a_{n+1}-(k-1)ba_{n}-2bna_{n}-lM_{n}^{\prime }
\]

\begin{equation}
+bM_{n-1}^{\prime }-T_{n}+2EM_{n-1}+b^{2}a_{n-1}-2BM_{n-2}=0,
\end{equation}
with the radial wave functions are still given in Eq.(15).

\subsection{Cornell Potential}

Now we investigate a confinement potential consisting of an attractive
Coulomb term and a confining linear potential used for calculation of
quarkonium $(q\overline{q})$ bound-state masses [20,21]. With the set of
parameters $V_{1}=A=4\alpha _{s}/3,$ $V_{2}=B,$ $V_{3}=C,$ $\alpha =$ $\beta
=1,$ then the recurrence relation (14) becomes

\[
n(n+1)a_{n+1}+(k-1)(n+1)a_{n+1}-(k-1)ba_{n}-2bna_{n}-lM_{n}^{\prime
}+bM_{n-1}^{\prime }-T_{n}
\]

\begin{equation}
+2EM_{n-1}+b^{2}a_{n-1}+2AM_{n}-2BM_{n-2}-2CM_{n-1}=0,
\end{equation}
with the radial wave functions are given in Eq.(15).

Let us investigate the last case. The recurrence relation (19) implies

\begin{equation}
a_{1}=\left[ b+\frac{lb_{0}^{\prime }-2Am_{0}}{k-1}\right] a_{0},
\end{equation}

\begin{equation}
a_{2}=\frac{\left[ (k+1)b+(l+1)b_{0}^{\prime }-2Am_{0}\right] }{2k}a_{1}+%
\frac{\left[ lb_{1}^{\prime }-bb_{0}^{\prime }-2Ab_{1}+2Cm_{0}\right] }{2k}%
a_{0},
\end{equation}

\[
a_{3}=\frac{\left[ (k+3)b+(l+2)b_{0}^{\prime }-2Am_{0}\right] }{3(k+1)}a_{2}+%
\frac{\left[ (l+1)b_{1}^{\prime }-bb_{0}^{\prime }-2Ab_{1}+2Cm_{0}\right] }{%
3(k+1)}a_{1}
\]

\begin{equation}
+\frac{\left[ lb_{2}^{\prime }-bb_{1}^{\prime }+2(C-E)b_{1}-2Ab_{2}+2Bm_{0}%
\right] }{3(k+1)}a_{0}.
\end{equation}

\section{The Analytical Series Solutions for $m=m_{0}e^{-\protect\lambda r}$}

\label{TAS}We choose one simple example considered recently by Ref.[18].
Assuming a particle with an exponentially decaying position-dependent
effective mass $m(r)=m_{0}e^{-\lambda r},$ $\lambda >0$ [18]. This form is
taken on the base that the position-dependent mass must be convergent when $%
r\rightarrow \infty .$ We try the series expansion

\begin{equation}
m(r)=\sum_{\nu =0}^{\infty }m_{\nu }r^{\nu },\text{ \ \ }m_{0}=1,
\end{equation}
and

\begin{equation}
\frac{m^{\prime }}{m}=-\lambda .
\end{equation}
Thus, for Cornell potential, Eq.(5) gives the following recursion relation

\[
\left[ (k-1)n+n(n-1)\right] a_{n}-\left[ b(k-1)+(2b-\lambda )(n-1)-l\lambda %
\right] a_{n-1}
\]

\begin{equation}
+b(b-\lambda
)a_{n-2}+2Em_{n-2}a_{n-2}+2Am_{n-1}a_{n-1}-2Bm_{n-3}a_{n-3}-2Cm_{n-2}a_{n-2}=0,
\end{equation}
which implies

\begin{equation}
a_{1}=\left[ b-\frac{l\lambda +2Am_{0}}{k-1}\right] a_{0},
\end{equation}

\begin{equation}
a_{2}=\frac{\left[ (k+1)b-\lambda (l+1)-2Am_{0}\right] }{2k}a_{1}+\frac{%
\left[ 2Cm_{0}+\lambda b-2Am_{1}\right] }{2k}a_{0},
\end{equation}

\[
a_{3}=\frac{\left[ (k+3)b-(l+2)\lambda -2Am_{0}\right] }{3(k+1)}a_{2}+\frac{%
\left[ \lambda b+2Cm_{0}-2Am_{1}\right] }{3(k+1)}a_{1}
\]

\begin{equation}
+\frac{\left[ 2Bm_{0}+2(C-E)m_{1}-2Am_{2}\right] }{3(k+1)}a_{0}.
\end{equation}
Finally, for another application of the model in case of a constant particle
mass $m=m_{0}.$ We find the Coulomb's wave functions.in three-dimensional
space, $N=3,$ easily through the relations (26)-(28). We present the
following results  [24]

\begin{equation}
a_{1}=-\frac{2Am_{0}}{(n+l+1)}\frac{n}{(2l+2)}a_{0},
\end{equation}

\begin{equation}
a_{2}=\left( -\frac{2Am_{0}}{n+l+1}\right) ^{2}\frac{n(n-1)}{2!(2l+3)(2l+2)}%
a_{0},
\end{equation}

\begin{equation}
a_{3}=\left( -\frac{2Am_{0}}{n+l+1}\right) ^{3}\frac{n(n-1)(n-2)}{%
3!(2l+4)(2l+3)(2l+2)}a_{0},
\end{equation}
and from which the wave functions is given by

\begin{equation}
R_{n,l}(r)=r^{l+1}e^{-\frac{Am_{0}}{(n+l+1)}r}\sum_{i=0}^{\infty
}(-1)^{i}\left( \frac{2Am_{0}}{n+l+1}\right) ^{i}\frac{(2l+1)!n!}{%
(2l+1+i)!(n-i)!i!}a_{0},
\end{equation}
with a proper normalization

\begin{equation}
a_{0}\approx N_{n,l}^{(C)}=\left[ \frac{2Am_{0}}{n+l+1}\right] ^{l+1}\frac{1%
}{(n+l+1)}\sqrt{\frac{Am_{0}n!}{(n+2l+1)!}},\text{ \ }n=0,1,2,\cdots .
\end{equation}
On the other hand, considering a particle with an exponentially decaying
position-dependent effective mass $m(r)=m_{0}e^{-\lambda r},$ the relations
(26)-(28) read [24]

\begin{equation}
a_{1}=\frac{Am_{0}}{(l+1)}\left[ \frac{l+1}{n+l+1}-\frac{l\lambda }{2m_{0}A}%
-1\right] a_{0},
\end{equation}

\begin{equation}
a_{2}=\frac{Am_{0}}{(2l+3)}\left\{ \left[ \frac{l+2}{n+l+1}-\frac{%
(l+1)\lambda }{2Am_{0}}-1\right] a_{1}+\left[ \frac{\lambda }{2(n+l+1)}-%
\frac{m_{1}}{m_{0}}\right] a_{0}\right\} ,
\end{equation}

\[
a_{3}=\frac{Am_{0}}{3(l+2)}\left\{ \left[ \frac{l+3}{n+l+1}-\frac{%
(l+2)\lambda }{2Am_{0}}-1\right] a_{2}+\left[ \frac{\lambda }{2(n+l+1)}-%
\frac{m_{1}}{m_{0}}\right] a_{1}\right.
\]
\begin{equation}
\left. -\left[ \frac{Am_{1}}{2(n+l+1)}-\frac{m_{2}}{m_{0}}\right]
a_{0}\right\} ,
\end{equation}
and from which the Coulomb's wave functions

\begin{equation}
R_{n,l}(r)=r^{l+1}e^{-\frac{Am_{0}}{(n+l+1)}r}\left[
a_{0}+a_{1}r+a_{2}r^{2}+a_{3}r^{3}+\cdots \right] .
\end{equation}
where $a_{i},$ with $i=0,1,2,3,$ are given by means of Eqs.(33)-(36).

\section{Concluding Remarks}

\label{CR}We have carried out the series solutions of
position-dependent mass $N$-dimensional radial Schr\"{o}dinger
equation for a general class of static potentials which are mostly
used in the calculations of quarkonium bound-state masses. Some
interesting results including the Coulomb, linear, harmonic, and
Cornell potentials are also considered. In particular, the Cornell
case was investigated and the series solution for the $N-$
dimensional radial Schr\"{o}dinger equation with an exponentialy
decaying position-dependent effective mass $m(r)=m_{0}e^{-\lambda
r}$, are solved analytically. Contrary to most of the approximation
methods which are valid when the parameters are small, our results
obtained by this approximation method are also valid with large
values of the parameters. These analytical calculations can be
applied to some related fields of physics such as quarkonium systems
in three dimensional space, i.e., $N=3.$ The work presented through
the two examples discussed above would appear to be of interest in
the ground that it offers an explicit formulas used to construct the
wave functions for a central potential [20,21,22,23,24].

\acknowledgments S.M. Ikhdair wishes to dedicate this work to his
family for their love and assistance.his research was partially
supported by the Scientific and Technical Research Council of
Turkey.

\bigskip

\end{document}